# Thermodynamics of Evolution and the Origin of Life

Vitaly Vanchurin[1,2,*], Yuri I. Wolf[1], Eugene V. Koonin[1,*], Mikhail I. Katsnelson[3,*]

[1]National Center for Biotechnology Information, National Library of Medicine, Bethesda, MD 20894;
[2]Duluth Institute for Advanced Study, Duluth, Minnesota, 55804, USA;
[3]Radboud University, Institute for Molecules and Materials, Nijmegen, 6525AJ, Netherlands;

*For correspondence: vitaly.vanchurin@gmail.com, m.katsnelson@science.ru.nl, koonin@ncbi.nlm.nih.gov

**Abstract**

We outline a phenomenological theory of evolution and origin of life by combining the formalism of classical thermodynamics with a statistical description of learning. The maximum entropy principle constrained by the requirement for minimization of the loss function is employed to derive a canonical ensemble of organisms (population), the corresponding partition function (macroscopic counterpart of fitness) and free energy (macroscopic counterpart of additive fitness). We further define the biological counterparts of temperature (biological temperature) as the measure of stochasticity of the evolutionary process and of chemical potential (evolutionary potential) as the amount of evolutionary work required to add a new trainable variable (such as an additional gene) to the evolving system. We then develop a phenomenological approach to the description of evolution, which involves modeling the grand potential as a function of the biological temperature and evolutionary potential. We demonstrate how this phenomenological approach can be used to study the "ideal mutation" model of evolution and its generalizations. Finally, we show that, within this thermodynamics framework, major transitions in evolution, such as the transition from an ensemble of molecules to an ensemble of organisms, that is, the origin of life, can be modeled as a special case of bona fide physical phase transitions that are associated with the emergence of a new type of grand canonical ensemble and the corresponding new level of description.

**Significance statement**

We employ the conceptual apparatus of thermodynamics to develop a phenomenological theory of evolution and of the origin of life that incorporates both equilibrium and non-equilibrium evolutionary processes within a mathematical framework of the theory of learning. The threefold correspondence is traced between the fundamental quantities of thermodynamics, the theory of learning and the theory of evolution. Under this theory, major transitions in evolution, including the origin of life, represent specific types of physical phase transitions.

## 1. Introduction

Classical thermodynamics is probably the best example of the efficiency of a purely phenomenological approach for the study of an enormously broad range of physical and chemical phenomena (1, 2). According to Einstein, "It is the only physical theory of universal content, which I am convinced, that within the framework of applicability of its basic concepts will never be overthrown" (3). Indeed, the basic laws of thermodynamics were established at a time when the atomistic theory of matter was only in its infancy, and even the existence of atoms has not yet been demonstrated unequivocally. Nevertheless, these laws remained untouched by all subsequent developments in physics, with the important qualifier "within the framework of applicability of its basic concepts". This framework of applicability is known as the "thermodynamic limit", the limit of a large number of particles when fluctuations are assumed to be small (4). Moreover, the concept of entropy that is central to thermodynamics was further generalized to become the cornerstone of information theory (Shannon's entropy (5)) and is currently considered to be one of the most important concepts in all of science, reaching far beyond physics (6).

It is therefore no surprise that many attempts have been made to apply concepts of thermodynamics to problems of biology, especially, to population genetics and the theory of evolution. The basic idea is straightforward: evolving populations of organisms fall within the domain of applicability of thermodynamics inasmuch as a population consists of a number of organisms sufficiently large for the predictable collective effects to dominate over unpredictable life histories of individual (where organisms are analogous to particles and their individual histories are analogous to thermal fluctuations). To our knowledge, this analogy was clearly formulated for the first time by Ronald Fisher, the principal founder of theoretical population genetics (7). Subsequently, extended efforts aimed at establishing detailed mapping between the principal quantities analyzed by thermodynamics, such as entropy, temperature and free energy, and those central to population genetics, such as effective population size and fitness, were undertaken by Sella and Hirsh (8), and elaborated by Barton and Coe (9). The parallel is indeed clear: the smaller the population size the stronger are the effects of random processes (genetic drift), which in physics associates naturally with temperature increase. It should be noted, however, that this crucial observation was predicated on the specific model of independent, relatively rare mutations (low mutation limit) in a constant environment or the so-called "ideal mutation" model. Among other attempts to conceptualize the relationships between thermodynamics and evolution, of special note is the work of Frank (10, 11) on applications of the maximum entropy principle, according to which the distribution of any quantity in a large ensemble of entities tends to the highest entropy distribution subject to the relevant constraints (6). The nature of such constraints, as we shall see, is the major object of enquiry in the study of evolution from the perspective of thermodynamics.

The deep parallels notwithstanding, the conceptual framework of classical thermodynamics is insufficient for an adequate description of evolving systems capable of learning. In such systems, the entropy increase due to physical and chemical processes in the environment under the second law of thermodynamics competes with the entropy decrease due to the second law of learning (12). Indeed, learning, by definition, decreases the uncertainty in knowledge and thus should result in entropy decrease. In the accompanying paper, we describe deep, multifaceted connections between learning and evolution, and outline a theory of evolution as learning (13). In particular, this theory incorporates a theoretical description of major transitions in evolution (MTE) (14, 15) and multilevel selection (16-19), two fundamental evolutionary phenomena that so far have not been fully incorporated into the theory of evolution.

Here, we make the next logical step towards a formal description of biological evolution as learning. Our main goal is to develop a macroscopic, phenomenological description of evolution in the spirit of classical thermodynamics, under the assumption that not only the number of degrees of freedom, but also the number of the learning subsystems (organisms or populations) is large. These conditions correspond to the thermodynamic limit in statistical mechanics, where the statistical description is accurate.

The paper is organized as follows. In Sec. 2 we apply the maximum entropy principle to derive a canonical ensemble of organisms and to define relevant macroscopic quantities, such as partition function and free energy. In Sec. 3 we discuss the first and second laws of learning, and their relations to the first and second laws of thermodynamics, in the context of biological evolution. In Sec. 4 we develop a phenomenological approach to evolution and define relevant thermodynamic potentials (such as average loss function, free energy, grand potential) and thermodynamic parameters (such biological temperature and evolutionary potential). In Sec. 5 we apply this phenomenological approach to analyze evolutionary dynamics of the "ideal mutations" model previously analyzed by Hirsch and Sella [7]. In Sec. 6 we demonstrate how the phenomenological description can be generalized to study more complex systems in the context of the "ideal gas" model. In Sec. 7 we apply the phenomenological description to model MTE, and in particular, the origin of life as a phase transition from an ideal gas of molecules to an ideal gas of organisms. Finally, in Sec. 8, we summarize the main facets of our phenomenological theory of evolution and discuss its general implications.

## 2. Maximum entropy principle applied to learning and evolution

To build the vocabulary of evolutionary thermodynamics (Table 1), we proceed step by step. The natural first concept to introduce is entropy, $S$, which is universally applicable beyond physics, thanks to the information representation of entropy (5). The relevance of entropy in general and the maximum entropy principle (6) for problems of population dynamics and evolution has been addressed previously, in particular, by Frank (20, 21), and we adopt this principle as our starting point. The maximum entropy principle states that the probability distribution in a large ensemble of variables must be such that the Shannon (or Boltzmann) entropy is maximized subject to the relevant constraints. This principle is applicable to an extremely broad variety of processes, but as shown below, is insufficient for an adequate description of learning and evolutionary dynamics, and should be combined with the opposite principle of minimization of entropy due to the learning process, or the second law of learning (see Sec. 3 and (12)). Our presentation in this section could appear over-simplified, but we find this approach essential to formulate as explicitly and as generally as possible all the basic assumptions underlying thermodynamics of learning and evolution.

**Table 1. Corresponding quantities in thermodynamics, machine learning and evolutionary biology**[a]

|  | **Thermodynamics** | **Machine learning** | **Evolutionary biology** |
|---|---|---|---|
| **x** | Microscopic physical degrees of freedom | Variables describing training dataset (non-trainable variables) | Variables describing environment |
| **q** | Generalized coordinates (e.g. volume) | Weight matrix and bias vector (trainable variables) | Trainable variables (genotype, phenotype) |
| $H(\mathbf{x}, \mathbf{q})$ | Energy | Loss function | Additive fitness, $H(x,q)=-T\log\phi(q)$ |
| $S(\mathbf{q})$ | Entropy of physical system | Entropy of non-trainable variables | Entropy of environment |
| $U(\mathbf{q})$ | Internal energy | Average loss function | Average additive fitness |
| $\mathcal{Z}(T, \mathbf{q})$ | Partition function | Partition function | Macroscopic fitness |
| $F(T, \mathbf{q})$ | Helmholtz free energy | Free energy | Adaptive potential (macroscopic additive fitness) |
| $\Omega(T, \mu)$ | Grand potential, $\Omega_p(\mathcal{T}, \mathcal{M})$ | Grand potential | Grand potential, $\Omega_b(T, \mu)$ |
| $T$ or $\mathcal{T}$ | Physical temperature, $\mathcal{T}$ | Temperature | Biological temperature, $T$ |
| $\mu$ or $\mathcal{M}$ | Chemical potential, $\mathcal{M}$ | Absent in conventional machine learning | Evolutionary potential, $\mu$ |
| $N_e$ or $N$ | Number of molecules, $N$ | Number of neurons, $N$ | Effective population size, $N_e$ |
| $K$ | Absent in conventional physics | Number of trainable variables | Number of adaptable variables |

[a]For further details, see text and refs. (12, 13).

The crucial step in treating evolution as learning is the separation of variables into trainable and non-trainable ones (13). The trainable variables are subject to evolution by natural selection, and therefore, should be related, directly or indirectly, to the replication processes, whereas non-trainable variables initially characterize the environment, which determines the criteria of selection. As an obvious example, chemical and physical parameters of the substances that serve as food for organisms are non-trainable variables, whereas the biochemical characteristics of proteins involved in the consumption and utilize the food molecules as building blocks and energy source are trainable variables.

Consider an arbitrary learning system described by trainable variables **q** and non-trainable variables **x**, such that non-trainable variables undergo stochastic dynamics, and trainable variables undergo learning dynamics. In the limit when the non-trainable variables $\mathbf{x}$ have already equilibrated, but the trainable variables $\mathbf{q}$ are still in the process of learning, the conditional probability distribution $p(\mathbf{x}|\mathbf{q})$ over non-trainable variables $\mathbf{x}$ can be obtained from the maximum entropy principle whereby Shannon (or Boltzmann) entropy

$$S = -\int d^N x \, p(\mathbf{x}|\mathbf{q}) \log p(\mathbf{x}|\mathbf{q}) \quad (2.1)$$

is maximized subject to the appropriate constraints on the system, such as average loss

$$\int d^N x \, H(\mathbf{x}, \mathbf{q}) p(\mathbf{x}|\mathbf{q}) = U(\mathbf{q}) \quad (2.2)$$

and normalization condition

$$\int d^N x \, p(\mathbf{x}|\mathbf{q}) = 1. \quad (2.3)$$

Its simplicity notwithstanding, the condition (2.2) is crucial. This condition means, first, that learning can be mathematically described as minimization of some function $U(\mathbf{q})$ of trainable variables only, and second, that this function can be represented as the average of some function $H(\mathbf{x}, \mathbf{q})$ of both trainable, $\mathbf{q}$, and non-trainable, $\mathbf{x}$, variables over the space of the latter. Equation (2.2) is not an assumption, but rather follows from the interpretation of the function $p(\mathbf{x}|\mathbf{q})$ as the probability density over non-trainable variables, $\mathbf{x}$, for a given set of trainable ones, $\mathbf{q}$. This condition is quite general and can be used to study, for example, selection of the shapes of crystals (such as snowflakes), in which case $H(\mathbf{x}, \mathbf{q})$ is simply free energy density.

In the context of biology, $U(\mathbf{q})$ is expected to be a monotonically increasing function of Malthusian fitness $\varphi(\mathbf{q})$, that is, reproduction rate (assuming a constant environment); a specific choice of this function will be motivated below (2.9). However, this connection cannot be taken as a definition of the loss function. In a learning process, loss function can be any measure of ignorance, that is, inability of an organism to recognize the relevant features of the environment and to predict its behavior. According to Sir Francis Bacon's famous motto, *scientia potentia est*: better knowledge and hence improved ability to predict the environment increases chances of an organism's survival and reproduction. However, derivation of the Malthusian fitness from the properties of the learning system requires a detailed microscopic theory such as that outlined in the accompanying paper (13). Here, we look instead at the theory from a macroscopic perspective by developing a phenomenological description of evolution.

We postulate that a system under consideration obeys the maximum entropy principle, but is also learning or evolving by minimizing the average loss function $U(\mathbf{q})$ (2.2). The corresponding maximum entropy distribution can be calculated using the method of Lagrange multipliers, that is, by solving the following variational problem

$$\frac{\delta \left(S - \beta \left(\int d^N y\, H(\mathbf{y}, \mathbf{q})p(\mathbf{y}|\mathbf{q}) - U\right) - \nu \left(\int d^N y\, p(\mathbf{y}|\mathbf{q}) - 1\right)\right)}{\delta p(\mathbf{x}|\mathbf{q})} = 0 \quad (2.4)$$

where $\beta$ and $\nu$ are the Lagrange multipliers which impose, respectively, the constraints (2.2) and (2.3). The solution of (2.4) is the Boltzmann (or Gibbs) distribution

$$-\log p(\mathbf{x}|\mathbf{q}) - 1 - \beta H(\mathbf{x}, \mathbf{q}) - \nu = 0$$

$$p(\mathbf{x}|\mathbf{q}) = \exp(-\beta H(\mathbf{x}) - 1 - \nu) = \frac{\exp(-\beta H(\mathbf{x}, \mathbf{q}))}{\mathcal{Z}(\beta, \mathbf{q})}, \quad (2.5)$$

where

$$\mathcal{Z}(\beta, \mathbf{q}) = \exp(1 + \nu) = \int d^N x\, \exp\left(-\beta H(\mathbf{x}, \mathbf{q})\right) = \int d^N x\, \varphi(\mathbf{x}, \mathbf{q}) \quad (2.6)$$

is the partition function ($\mathcal{Z}$ stands for German *Zustandssumme*, sum over states).

Formally, the partition function $\mathcal{Z}(\beta, \mathbf{q})$ is simply a normalization constant in equation (2.5), but its dependence on $\beta$ and $\mathbf{q}$ contains a wealth of information about the learning system and its environment. For example, if the partition function is known, then, the average loss can be easily calculated by simple differentiation

$$U(\mathbf{q}) = \frac{\int d^N x\, H(\mathbf{x}, \mathbf{q}) \exp\left(-\beta H(\mathbf{x}, \mathbf{q})\right)}{\int d^N x\, \exp\left(-\beta H(\mathbf{x}, \mathbf{q})\right)} = -\frac{\partial}{\partial \beta} \log \mathcal{Z}(\beta, \mathbf{q}) = \frac{\partial}{\partial \beta}\left(\beta F(\beta, \mathbf{q})\right) \quad (2.7)$$

where the biological equivalent of free energy is defined as

$$F \equiv -T \log(\mathcal{Z}) = -\beta^{-1} \log(\mathcal{Z}) = U - TS \quad (2.8)$$

and the biological equivalent of temperature is $T = \beta^{-1}$. Biological temperature is yet another key term in our vocabulary (Table 1), after entropy, which emerges as the inverse of the Lagrange multiplier $\beta$ that imposes a constraint on the average loss function (2.2), that is, defines the extent of stochasticity of the process of evolution. Roughly, free energy $F$ is the macroscopic counterpart of the loss function $H$ or additive fitness (13) whereas, as shown below, partition function $\mathcal{Z}$ is the macroscopic counterpart of Malthusian fitness,

$$\varphi \equiv \exp(-\beta H(\mathbf{x}, \mathbf{q})) \quad (2.9)$$

The relation between the loss function and fitness is discussed in the accompanying paper (13) and in Sec. 5. In biological terms, $\mathcal{Z}$ represents macroscopic fitness or the sum over all possible fitness values for a given organism, that is, over all genome sequences that are compatible with survival in a given environment, whereas $F$ represents the adaptation potential of the organism.

## 3. Thermodynamics of learning

In the rest of this analysis, we follow the previously developed approach to the thermodynamics of learning

(12). Here, the key difference from conventional thermodynamics is that learning decreases the uncertainty in our knowledge of the training dataset (or in our case of the environment) and therefore results in the entropy decrease. Close to the learning equilibrium, this decrease compensates exactly for the thermodynamic entropy increase and such dynamics is formally described by a time-reversible Schrödinger-like equation (12, 22). An important consequence is that, whereas in conventional thermodynamics, the equilibrium corresponds to the minimum of the thermodynamic potential over all variables, in a learning equilibrium, the free energy $F(\mathbf{q})$ can either be minimized or maximized with respect to the trainable variables $\mathbf{q}$. If for a particular trainable variable, the entropy decrease due to learning is negligible, then, the free energy is minimized, as in conventional thermodynamics, but if the entropy decrease dominates the dynamics, then, the free energy is maximized. Using the terminology introduced in the accompanying paper (13), we will call the variables of the first type *neutral* $\mathbf{q}^{(n)}$, and those of the second type *adaptable* or *active* variables $\mathbf{q}^{(a)}$. There is also a third type of variables that are (effectively) *constant* or *core* variables $\mathbf{q}^{(c)}$, that is, those that have already been well trained. The term neutral means that changing the values of these variables does not affect the essential properties of the system, such as its loss function or fitness, corresponding to the regime of neutral evolution. The adaptable variables comprise the bulk of the material for evolution. The core variables are most important for optimization (that is, for survival) and thus are quickly trained to their optimal values and remain more or less constant during the further process of learning (evolution). The equilibrium state corresponds to a saddle point on the free energy landscape (viewed as a function of trainable variables $\mathbf{q}$), in agreement with both the first law of thermodynamics and the first law of learning (12):

*First law:* *the change in loss/energy equals the heat added to the learning/thermodynamic system minus the work done by the system*

$$dU = TdS - \mathbf{Q} \cdot d\mathbf{q} \quad (3.1)$$

*where $T$ is temperature, $S$ is entropy and $\mathbf{Q}$ is the learning/generalized force for the trainable/external variables $\mathbf{q}$.*

In the context of evolution, the first term in Eq. (3.1) represents the stochastic aspects of the dynamics, whereas the second term represents adaptation (learning, work). If the state of the entire learning system is such that the learning dynamics is subdominant to the stochastic dynamics, then, the total entropy will increase (as is the case in regular, closed physical systems, under the second law of thermodynamics), but if learning dominates, then, entropy will decrease as is the case in learning systems, under the second law of learning (12):

*Second law: The total entropy of a thermodynamic system does not decrease and remains constant in the thermodynamic equilibrium, but the total entropy of a learning system does not increase and remains constant in the learning equilibrium.*

If the stochastic entropy production and the decrease in entropy due to learning cancel out each other, then, the overall entropy of the system remains constant and the system is in the state of learning equilibrium (see (12, 22, 23) for discussion of different aspects of the equilibrium states.) This second law, when applied to biological processes, specifies and formalizes Schrödinger's idea of life as a "negentropic" phenomenon (24). Indeed, the state of learning equilibrium is the fundamental stationary state of biological systems.

On longer time-scales, when $\mathbf{q}^{(c)}$ remains fixed, but all other variables (i.e. $\mathbf{q}^{(a)}$, $\mathbf{q}^{(n)}$ and **x**) have equilibrated, the adaptable variables $\mathbf{q}^{(a)}$ can transform into neutral ones $\mathbf{q}^{(n)}$ and vice versa, neutral variables can become adaptable ones (13). In terms of statistical mechanics, such transformations can be described by generalizing the canonical ensemble with the fixed number of particles (that is, in our context, fixed number of variables relevant for training) to a grand canonical ensemble where the number of variables can fluctuate (2). For neural networks, such fluctuations correspond to recruiting additional neurons from the environment or excluding neurons from the learning process. On a phenomenological level, these transformations can be described as finite shifts in the loss function, $U \to U \pm \mu$. In conventional thermodynamics, when dealing with ensembles of particles, $\mu$ is known as chemical potential, but in the context of biological evolution, we shall refer to $\mu$ as *evolutionary* potential, another key term in our vocabulary of evolutionary thermodynamics and learning (Table 1). In thermodynamics, chemical potential describes how much energy is required to move a particle from one phase to another (for example, moving one water molecule from liquid to gaseous phase during water evaporation). Analogously, the evolutionary potential corresponds to the amount of evolutionary work (expressed, for example, as the number of mutations) or the magnitude of the change in the loss function is associated with the addition or removal of a single adaptable variable to or from the learning dynamics, that is, how much work does it take to make a non-adaptable variable adaptable, or vice versa.

The concept of evolutionary potential, $\mu$, has multiple important connotates in evolutionary biology. Indeed, it is recognized that networks of nearly neutral mutations, and more broadly, non-functional genetic material ("junk DNA") that dominates the genomes of complex eukaryotes represents the reservoir of potential adaptations (25-28) making the evolutionary cost of adding a new adaptable variable low which corresponds to small $\mu$. Genomes of prokaryotes are far more tightly constrained by purifying selection and thus contain little junk DNA (29, 30); put another way, the evolutionary potential $\mu$ associated with such neutral genetic sequences is high in prokaryotes. However, this comparative evolutionary rigidity of prokaryote genomes is compensated by the high rate of gene replacement (31), with vast pools of diverse DNA sequences (open pangenomes) available for acquisition of new genes, some of which can contribute to adaptation (32, 33). The cost of gene acquisition varies greatly among genes of different functional classes as captured in the genome plasticity parameter of genome evolution that effectively corresponds to the evolutionary potential introduced here (34). For many classes of genes in prokaryotes, the evolutionary potential $\mu$ is relatively lower, such that gene replacement represents the principal route of evolution in these life forms. In viruses, especially, those with RNA and single-stranded DNA genomes, the evolutionary potential associate with gene acquisition is prohibitively high, but this is compensated by high mutations rates (35, 36), that is, low evolutionary potential $\mu$ associated with extensive nearly neutral mutational networks, making these networks the main source of adaptation.

Treating the learning system as a grand-canonical ensemble, equation (3.1), which represents the first law of learning, can be rewritten as

$$dU = TdS + \mu dK \quad (3.2)$$

where $K$ is the number of adaptable variables. Equation (3.2) is more macroscopic than (3.1) in the sense that not only non-trainable variables, but also adaptable and neutral trainable variables are now described in terms of phenomenological, thermodynamic quantities. Roughly, the average loss associated with a single non-trainable or a single adaptable variable can be identified, respectively, with $T$ and $\mu$, and the total number of non-trainable and adaptable variables with, respectively, $S$ and $K$. This correspondence stems from the fact that $S$ and $K$ are extensive variables, whereas $T$ and $\mu$ are intensive ones, as in conventional thermodynamics.

To describe phase transitions, we have to consider the system moving from one learning equilibrium (that is, a saddle point on the free energy landscape) to another. In terms of the microscopic dynamics, such phase transitions can involve either transitions from not fully trained adaptable variables $q^{(a)}$ to fully trained ones $q^{(c)}$ or transitions between different learning equilibria described by different values of $q^{(c)}$. In biological terms, the latter variety of transitions corresponds to MTE, which involve emergence of new classes of slowly changing, near constant variables (13), whereas the former variety of smaller scale transitions corresponds to the fixation of beneficial mutations of all kinds, including capture of new genes (37), that is, adaptive evolution. In Sec. 7 we present a phenomenological description of MTE, in particular, the very first one, the origin of life, which involved the transition from an ensemble of molecules to an ensemble of organisms. First, however, we describe how such ensembles can be modeled phenomenologically.

## 4. Phenomenology of evolution

Consider an ensemble of organisms that differ from each other only by the values of adaptable variables $q^{(a)}$, whereas the effectively constant variables $q^{(c)}$ are the same for all organisms. The latter correspond to the set of core, essential genes that are responsible for the house-keeping functions of the organisms (38). Then, the ensemble can either represent a Bayesian (subjective) probability distribution over degrees of freedom of a single organism or a frequentist (objective) probability distribution over the entire population of organisms; the connections between these two approaches are addressed in detail in the classic work of Jaynes (6). In the limit of an infinite number of organisms, the two interpretations are indistinguishable, but in the context of actual biological evolution, the total number of organisms is only exponentially large

$$N_e \propto \exp(bK). \quad (4.1)$$

and is linked to the number of adaptable variables $K \propto \log(N_e)/b$ in a population of the given size $N_e$. Eq. (4.1) indicates that the effective number of variables (genes or sites in the genome) that are available for adaptation in a given population depends on the effective population size. In larger populations that are mostly immune to the effect of random genetic drift, more sites (genes) can be involved in adaptive evolution. In addition to the effective population size $N_e$, the number of adaptable variables depends on the coefficient $b$ that can be thought of as the measure of stochasticity caused by factors independent of the population size. The smaller $b$, the more genes can be involved in adaptation. In the biological context, this implies that the entire adaptive potential of the population is determined by mutations in a small fraction of the genome, which is indeed realistic. It has been shown that in prokaryotes effective population size estimated from the ratio of the rates of non-synonymous vs synonymous mutations (*dN/dS*), indeed, positively correlates with the number of genes in the genome, and presumably, with the number of genes that are available for adaptation (39-41).

To study the state of learning equilibrium for a grand canonical ensemble of organisms, it is convenient to express the average loss function as

$$U(S, K) = T(S, K)S + \mu(S, K)K \quad (4.2)$$

where the conjugate variables are, respectively, biological temperature

$$T \equiv \frac{\partial U}{\partial S}, \quad (4.3)$$

and evolutionary potential

$$\mu \equiv \frac{\partial U}{\partial K}. \quad (4.4)$$

Once again, biological temperature is a measure of disorder, that is, stochasticity in the evolutionary process, whereas evolutionary potential is the measure of adaptability. For a given phenomenological expression of the loss function (4.2), all other thermodynamic potentials, such as free energy $F(T, K)$ and grand potential $\Omega(T, \mu)$, can be obtained by switching to conjugate variables using Legendre transformations, i.e. $S \leftrightarrow T$, $K \leftrightarrow \mu$.

The difference between the grand canonical ensembles in physics and in evolutionary biology should be emphasized. In physics, the grand canonical ensemble is constructed by constraining the average number of particles (2). In contrast, for the evolutionary grand canonical ensemble, the constraint is imposed not on the number of organisms $N_e$ per se but rather on the number of adaptable variables in organisms of a given species $K \propto \log(N_e)$, which depends on the effective population size. This key statement implies that, in our approach, the primary agency of evolution (adaptation, selection, or learning) is identified with individual genes rather than with genomes and organisms (42). Only a relatively small number of genes represent adaptable variables, that is, are subject to selection at any given time, in accordance with the classical results of population genetics (43). However, as discussed in the accompanying paper (13), our theoretical framework extends to multiple levels of biological organization and is centered around the concept of multilevel selection such that higher-level units of selection are identified with ensembles of genes or whole genomes (organisms). Then, organisms can be treated as trainable variables (units of selection) and populations as statistical ensembles. The change in the constraint from $N_e$ to $K \propto \log(N_e)$ is similar to changing the ensemble with annealed disorder to one with quenched disorder in statistical physics (44). Indeed, in the case of annealed (thermal) disorder, we sum up (average) over a disorder partition function, whereas for quenched disorder, we average the logarithm of the partition function, that is, free energy.

## 5. Ideal mutation model

In this section, we demonstrate how the phenomenological approach developed in the previous sections can be applied to model biological evolution in the thermodynamic limit, that is, when both the number of organisms, $N_e$, and the number of active degrees of freedom, $K \propto \log(N_e)$, are sufficiently large. In such a limit, the average loss function contains all the relevant information on the learning system in equilibrium, which can be derived from a theoretical model, such as the one developed in the accompanying paper (13) using the mathematical framework of neural networks, or a phenomenological model (such as the one developed in the previous section), or reconstructed from observations or numerical simulations. In this section, we adopt a phenomenological approach to model the average loss function of a population of non-interacting organisms (that is, selection only affects individuals), and in the following section, we construct a more general phenomenological model, which will also be relevant for the analysis of MTE in Sec. 7.

Consider a population of organisms described by their genotypes $\mathbf{q}_1, \cdots, \mathbf{q}_{N_e}$. There are rare mutations (on time-scales $\sim \tau$) from one genotype to another, that are either quickly fixed or eliminated from the population (on shorter time-scales $\ll \tau$), but the total number of organisms $N_e$ remains fixed. In addition, we assume that the system is observed for a long period of time $\gg \tau$ so that it has reached a learning equilibrium (that is, an evolutionarily stable configuration). In this simple model, all organisms share the same $\mathbf{q}^{(c)}$ whereas all other variables have already equilibrated, but their effect on the loss function depends on the type of the variable, that is, $\mathbf{q}^{(a)}$ vs. $\mathbf{q}^{(n)}$ vs. $\mathbf{x}$. In particular, the trainable variables of individual organisms $\mathbf{q}_n$'s evolve in such a way that entropy is minimized on short time-scales $\ll \tau$ due to fixation of beneficial mutations but maximized on long time-scales $\gg \tau$ due to equilibration, that is, exploration of the entire nearly neutral mutational network (45, 46). Thus, the same variables evolve towards maximizing free energy on short time scales, but towards minimizing free energy on longer time-scales. This evolutionary trajectory is similar to the phenomenon of broken ergodicity in condensed matter systems, where the short-time and ensemble (or long-time) averages can differ. The prototype of nonergodic systems in physics are (spin) glasses (47-49). The glass-like character of evolutionary phenomena was qualitatively examined previously (50, 51). Nonergodicity unavoidably involves frustrations that emerge from competing interactions (52), and such frustrations are thought to be a major driving force of biological evolution (50). In terms of the model we discuss here, the most fundamental frustration that appear to be central to evolution is caused by the competing trends of the conventional thermodynamic entropy growth and entropy decrease due to learning.

The fixation of mutations on short time-scales $\ll \tau$ implies that over most of the duration of evolution all organisms have the same genotype $\mathbf{q}_1 = \cdots = \mathbf{q}_{N_e} = \mathbf{q}$ (with some neutral variance), whereas the equilibration on the longer time-scales $\gg \tau$ implies that the marginal distribution of genotypes is given by the maximum entropy principle as discussed in Sec. 2, that is,

$$p(\mathbf{q}) \propto \int \prod_{n=1}^{N_e} d^N x_n \exp\left(-\beta \sum_{n=1}^{N_e} H(\mathbf{x}_n, \mathbf{q})\right) = \exp\left(-\beta F(\mathbf{q}) N_e\right) \quad (5.1)$$

where integration is taken over the states of the environment $\mathbf{x}_n$ for all organisms $n = 1 \ldots N_e$. This distribution was previously considered in the context of population dynamics (8), where $N_e$ was interpreted as the inverse temperature parameter. However, as pointed out in Sec. 2, in our framework, the inverse temperature $\beta$ is the Lagrange multiplier, which imposes a constraint on the average loss function (2.2). Moreover, in the context of the models considered by Sella and Hirsh (8), the distribution can also be expressed as

$$p(\mathbf{q}) \propto \mathcal{Z}(\mathbf{q})^{N_e} \quad (5.2)$$

where the partition function $\mathcal{Z}(\mathbf{q}) = \exp(-\beta F(\mathbf{q}))$ is the macroscopic counterpart of fitness $\varphi(\mathbf{x}, \mathbf{q})$ (see Eq. 2.6). Equation (5.2) implies that biological temperature has to be identified with the multiplication constant in (2.8), $T = \beta^{-1}$. Thus, this is the "ideal mutation" model, which allows us to establish a precise relation between the loss function and Malthusian fitness. Importantly, this relation holds only for the situation of multiple, non-interacting mutations (i.e. without epistasis).

The model of Sella and Hirsh (8), is actually the Kimura model of fixation of mutations in a finite population (53), which is based on the effect of mutations on Malthusian fitness (in the absence of epistasis). In population genetics, this model plays a role analogous to the role of the ideal gas model in statistical physics and thermodynamics (1). The ideal gas model ignores interactions between molecules in the gas, and the population genetics model similarly ignores epistasis, that is, interaction between mutations. This model necessitates that the loss function is identified with minus logarithm of Malthusian fitness (otherwise, the connection between these two quantities would be arbitrary, with the only restriction that one of them should be a monotonically decreasing function of the other). However, identification of $N_e$ with the inverse temperature $\beta$ (8) does not seem to be justified equally well. For the given environment, the probability of the state (5.1) depends only on the product of $N_e$ and $\beta$, that is, the parameter of the Gibbs distribution. This parameter is proportional to $N_e$, so that we could, in principle, choose the proportionality coefficient to be equal to 1 (or, more precisely, 1, 2 or 4 depending on genome ploidy and the type of reproduction), but only assuming that the properties of the environment are fixed. However, in the interest of generality, we keep the population size and the "biological temperature" separate, interpreting $\beta$ as an overall measure of the level of stochasticity in the evolutionary process including effects independent of the population size.

This key point merits further explanation. The smaller the population size the more important are evolutionary fluctuations, that is, genetic drift (54). In statistical physics, the amplitude of fluctuations increases with the temperature (2). Therefore, when the correspondence between evolutionary population genetics and thermodynamics is explored, it appears natural to identify effective population size with the inverse temperature (8, 9), which is justified inasmuch as sources of noise independent of the population size, such as changes in the environment, are disregarded. In statistical physics, the probability of a system leaving a local optimum at a given temperature exponentially depends on the number of particles in the system as compellingly illustrated by the phenomenon of superparamagnetism (55). For a small enough ferromagnetic particle, the total magnetic moment overcomes the anisotropy barrier and oscillates between the spin-up and spin-down directions, whereas in the thermodynamic limit, these oscillations are forbidden, which results in spontaneously broken symmetry (2). Thus, the probability of "drift" from one optimum to the other exponentially depends on the number of particles, and the identification of the latter with the effective population size appears natural. However, from a more general standpoint, effective population size is not the only parameter determining the probability of fluctuations, which is also affected by environmental factors. In particular, stochasticity increases dramatically under harsh conditions, due to stress-induced mutagenesis (56-58). Therefore, it appears preferable to represent biological temperature as a general measure of evolutionary stochasticity, to which effective population size is only one of important contributors.

Importantly, this simple model allows us to make concrete predictions for a fixed size population, where beneficial mutations are rare and quickly proceed to fixation. If such a system evolves from one equilibrium state (at temperature $T_1 = \beta_1^{-1}$, with the fitness distribution $\mathcal{Z}^{(1)}(\mathbf{q})$), to another equilibrium state at temperature $T_2 = \beta_2^{-1}$ and fitness distribution $\mathcal{Z}^{(2)}(\mathbf{q})$), then, according to (2.8), the ratios

$$\frac{\log \mathcal{Z}^{(1)}(\mathbf{q})}{\log \mathcal{Z}^{(2)}(\mathbf{q})} = \frac{\beta_1 F(\mathbf{q})}{\beta_2 F(\mathbf{q})} = \frac{\beta_1}{\beta_2} = \frac{T_2}{T_1} \quad (5.3)$$

are independent of $\mathbf{q}$, that is, are the same for all organisms in the ensemble, regardless of their fitness (again, under the key assumption of no epistasis). Then, Eq. (5.3) can be used to measure ratios between different biological temperatures and thus to define a temperature scale. Moreover, the equilibrium distribution (5.1) together with (4.1) enables us to express the average loss function

$$U(K) = \langle H(\mathbf{x}, \mathbf{q}) N_e \rangle \propto \langle H(\mathbf{x}, \mathbf{q}) \rangle \exp(bK), \quad (5.4)$$

where $\langle H(\mathbf{x}, \mathbf{q}) \rangle$ is the average loss function of individual organisms. According to Eq. (5.4), the average loss $U(S, K)$ scales exponentially with the number of adaptable degrees of freedom $K$, but the dependency on entropy is not yet explicit.

### 6. Ideal gas of organisms

For phenomenological modeling of evolution, it is essential to keep track not only of different organisms, but also of the entropy of the environment. On the microscopic level, the overall average loss function is an integral over all non-trainable variables of all organisms, but on a more macroscopic level, it can be viewed as a phenomenological function $U(S, K)$, the microscopic details of which are irrelevant. In principle, it should be possible to reconstruct the average loss function directly from experiment or simulation, but for the purpose of illustration, we first consider an analytical expression

$$U(S, K) = \langle H(\mathbf{x}, \mathbf{q}) \rangle N_e = aS^n \exp\left(\frac{b}{S} K\right) \quad (6.1)$$

where in addition to the exponential dependence on $K$, as in (5.4), we also specify the power law dependency on $S$. In particular, we assume that $\langle H(\mathbf{x}, \mathbf{q}) \rangle \propto S^n$, where $n > 0$ is a free parameter, that is, loss function is greater in an environment with a higher entropy. This factor models the effect of the environment on the loss function of individual organisms. In biological terms, this means that diverse and complex environments promote adaptive evolution. In addition, the coefficient $b$ in (5.4) is replaced with $b/S$ in (6.1), to model the effect of the environment on the effective number of active trainable variables. We have to emphasize that the model (6.1) is discussed here for the sole purpose of illustrating the way our formalism works. A realistic model can be built only through bioinformatic analysis of specific biological systems, which requires a major effort.

Thus, if a population of $N_e$ organisms is capable of learning the amount of information $S$ about the environment, then, the total number of adaptable trainable variables $K$ required for such learning scales linearly with $S$ and logarithmically with $N_e$,

$$K = b^{-1} S \log N_e. \quad (6.2)$$

The logarithmic dependence on $N_e$ is already present in (4.1) and in (5.4), but the dependence on $S$ is an addition introduced in the phenomenological model (6.1). Under this model, the number of adaptable variables $K$ is proportional to the entropy of the environment. Assuming $K$ is proportional also to the total number of genes in the genome, the dependencies in Eq. (6.2) are at least qualitatively supported by comparative analysis of microbial genomes. Indeed, bacteria that inhabit high entropy environments, such as soil, typically possess more genes than those that live in low entropy environments, for example, sea water (59). Furthermore, the number of genes in bacterial genomes increases with the estimated effective population size (39-41), which also can be interpreted as taking advantage of diverse, high entropy environments.

Given a phenomenological expression for the average loss function (6.1), the corresponding grand potential is given by the Legendre transformation,

$$\Omega(T,\mu) = (1-n)S(T)\frac{\mu}{b}, \quad (6.3)$$

where entropy should be expressed as a function of biological temperature and evolutionary potential,

$$T = \frac{\mu}{b}\left(n + \log\left(\frac{ab}{\mu}\right) + (n-1)\log(S)\right). \quad (6.4)$$

By solving (6.4) for $S$ and plugging into (6.3), we obtain the grand potential

$$\Omega(T,\mu) = -a(n-1)\left(\frac{\mu}{eb}\right)^{\frac{n}{n-1}} \exp\left(\frac{bT}{(n-1)\mu}\right). \quad (6.5)$$

We shall refer to the ensemble described by (6.5) as an "ideal gas" of organisms.

In principle, the grand potential can also be reconstructed phenomenologically, directly from numerical simulations or observations of time-series of the population size $N_e(t)$ and fitness distribution $\mathcal{Z}(\mathbf{q},t)$. Given such data, biological temperature $T$ can be calculated using (5.3) and the distributions $p_T(K) = p(\log(N_e))$ of the number of adaptable variables $K$ can be estimated for a given temperature $T$. Then, the grand potential is reconstructed from the cumulants $\kappa_n(T)$ of the distributions $p_T(K)$

$$\Omega(T,\mu) = -T\sum_{n=1}^{\infty}\frac{\kappa_n(T)}{n!}\left(\frac{\mu}{T}\right)^n \quad (6.6)$$

and the average loss function $U(S,K)$ is obtained by Legendre transformation from variables $(T,\mu)$ to $(S,K)$. Obviously, the phenomenological reconstruction of the thermodynamic potentials $\Omega(T,\mu)$ and $U(S,K)$ is feasible only if the evolving learning system can be observed over a long period of time, during which the system visits different equilibrium states at different temperatures. More realistically, the observation can be limited to either a fixed temperature $T$ or a fixed number of adaptable variables $K$, and then, the thermodynamic potentials would be reconstructed in the respective variables only, that is, in $K$ and $\mu$ or in $T$ and $S$.

### 7. Major transitions in evolution and the origin of life

In this section, we discuss the MTE, starting from the very first such transition, the origin of life. Under the definition of life adopted by NASA, natural selection is the quintessential trait of life. Here we assume that selection emerges from learning. which appears to be a far more general feature of the processes that occur on all scales in the universe (13, 60). Indeed, any statistical ensemble of molecules is governed by some optimization principle, which is equivalent to the standard requirement of minimization of the properly chosen potential in thermodynamics. Evolving populations of organisms similarly face an optimization problem, but at face value, the nature of the optimized potential is completely different. So what if anything is in common between thermodynamic free energy and Malthusian fitness? Here we give a specific answer to this question: the unifying feature is that, at any stage of the evolution or learning dynamics, the loss function is optimized. Thus, as also discussed in the accompanying paper (13), the origin of life is not equal to the origin of learning or selection. Instead, we associate the origin of life with a phase transition that gave rise to a distinct, highly efficient form of learning or a learning algorithm known as natural selection. Neither the nature of the statistical ensemble of molecules that preceded this phase transition nor that of the statistical ensemble of organisms that emerged from the phase transition (referred to as the Last Universal Cellular Ancestor, LUCA(61, 62)) are well understood, but at the phenomenological level, we can try to determine which statistical ensembles yield the most biologically plausible results.

The origin of life can be identified with a phase transition from an ideal gas of molecules that is often considered in the analysis of physical systems to an ideal gas of organisms that is discussed in the previous section. Then, during such a transition, the grand canonical ensemble of subsystems changes from being constrained by a fixed average number of subsystems (or molecules),

$$\langle N_e \rangle = \bar{N}_e \quad (7.1)$$

to being constrained by a fixed average number of adaptable variables associated with the subsystems (or organisms),

$$\langle K \rangle = \bar{K}. \quad (7.2)$$

Immediately before and immediately after the phase transition, we are dealing with the very same system, but the ensembles are described in terms of different ensembles of thermodynamic variables. Formally, it is possible to describe an organism by the coordinates of all atoms of which it is comprised, but this is not a particularly useful language (63). Atoms (and molecules) behave in a collective manner, that is, coherently, and therefore, the appropriate language to describe their behavior is the language of collective variables similar to, for example, the "dual boson" approach to many-body systems (64).

According to (4.1), the total number of organisms (population size) and the number of adaptable variables are related, $K \propto \log(N_e)$, but the choice of the constraint, (7.1) vs. (7.2), determines the choice of the statistical ensemble, which describes the state of the system. In particular, an ensemble of molecules can be described by the grand potential $\Omega_p(\mathcal{T}, \mathcal{M})$, where $\mathcal{T}$ is the physical temperature, $\mathcal{M}$ is the chemical potential, and an ensemble of biological subsystems can be described by the grand potential $\Omega_b(T, \mu)$, where, as before, $T$ is the biological temperature and $\mu$ is the evolutionary potential. Assuming that both ensembles can coexist at some critical temperatures $\mathcal{T}_c$ and $T_c$, the evolutionary phase transition will occur when

$$\Omega_p(\mathcal{T}_c, \mathcal{M}_c) = \Omega_b(T_c, \mu_c). \quad (7.3)$$

This condition is highly non-trivial because it implies that, at phase transition, both physical and biological potentials provide fundamentally different (or dual) descriptions of the exact same system, and all of the biological and physical quantities have different (or dual) interpretations. For example, the loss function is to be interpreted as energy in the physical description, but as additive fitness in the biological description (2.9).

An ideal gas of molecules is described by the grand potential

$$\Omega_p(\mathcal{T}, \mathcal{M}) \propto \mathcal{T}^\alpha \exp\left(\gamma \frac{\mathcal{M}}{\mathcal{T}}\right) \quad (7.4)$$

and an ideal gas of organisms is described by the grand potential (6.5),

$$\Omega_b(T, \mu) \propto \mu^c \exp\left(b \frac{T}{\mu}\right). \quad (7.5)$$

At higher temperatures, it is more efficient for the individual subsystems to remain independent of each other, $\Omega_p < \Omega_b$, but at lower temperatures, sharing of degrees of freedom between subsystems becomes beneficial such that $\Omega_b < \Omega_p$. Thus, a critical temperature exists that corresponds to the transition (7.3). The macroscopic quantities in (7.4) and (7.5) characterize the same system, but in terms of different (or dual) statistical ensembles, and so in general, only one of them would be relevant at any given time. However, in the immediate vicinity the phase transition (7.3), both phases can coexist and so the macroscopic quantities (i.e. $T$, $\mu$, $\mathcal{T}$ and $\mathcal{M}$) must all be related to each other. For example, consider the following relations (or dual mappings between physical and biological quantities)

$$T_c = \frac{\alpha}{b} \mu_c \log(\mathcal{T}_c) \quad (7.6)$$

and

$$\mathcal{M}_c = \frac{c}{\gamma} \mathcal{T}_c \log(\mu_c). \quad (7.7)$$

Plugging (7.6) and (7.7) into (7.4) gives

$$\Omega_p(\mathcal{T}_c, \mathcal{M}_c) \propto \mathcal{T}_c^\alpha e^{\gamma \mathcal{M}_c/\mathcal{T}_c} = \left(e^{\frac{bT_c}{\alpha \mu_c}}\right)^\alpha e^{c\log(\mu_c)} = e^{b\frac{T_c}{\mu_c}} \mu_c^c \propto \Omega_b(T_c, \mu_c), \quad (7.8)$$

which is in agreement with (7.3). The relations (7.6) and (7.7) were used here to illustrate the conditions under which the phase transition might occur, but it is also interesting to examine whether these relations actually make sense qualitatively. Equation (7.6) implies that energy/loss associated with learning dynamics, $T_c$, is logarithmically smaller compared to the energy/loss associated with stochastic dynamics, $\mathcal{T}_c$, but depends linearly on the energy/loss required to add a new adaptable variable to the learning system, that is the evolutionary potential $\mu_c$. This dependency makes sense because the learning dynamics is far more stringently constrained than the stochastic dynamics and its efficiency critically depends on the ability to engage new adaptable degrees of freedom. Equation (7.7) implies that the energy/loss, $\mathcal{M}_c$, that is required to incorporate an additional non-trainable variable into the evolving system is logarithmically smaller $\mu_c$, but depends linearly on the energy/loss, $\mathcal{T}_c$, associated with stochastic dynamics. This also makes sense because it is much easier to engage non-trainable degrees of freedom, and furthermore, the capacity of the system to do so depends on the physical temperature.

It appears that for the origin of life phase transition to occur, the learning system has to satisfy at least three conditions. The first one is the existence of effectively constant degrees of freedom, $\mathbf{q}^{(c)}$, which are the same in all subsystems. This condition is satisfied, for example, for an ensemble of molecules, the stability of which is a prerequisite of the evolutionary phase transitions, but it does not guarantee that the transition occurs. The second condition is the existence of adaptable or active variables, $\mathbf{q}^{(a)}$ that are shared by all subsystems, but their values can vary. These are the variables that undergo learning evolution and, according to the second law of learning, adjust their values to minimize entropy. Finally, for learning and evolution to be efficient, the third condition is the existence of neutral variables, $\mathbf{q}^{(n)}$, which can become adaptable variables as learning progresses. In the language of statistical ensembles, this is equivalent to switching from a canonical ensemble with a fixed number of adaptable variables to a grand canonical ensemble where the number of adaptable variables can vary.

There are clear biological connotates of these three conditions. In the accompanying paper (13), we identify the origin of life with the advent of natural selection which requires genomes that serve as instructions for the reproduction of organisms. Genes comprising the genomes are shared by the organisms in a population or community, forming expandable pangenome that can acquire new genes, some of which contribute to adaptation (32). In each prokaryote genome, about 10% of the genes are rapidly replaced, with the implication that they represent neutral variables that are subject to no or weak purifying selection and comprise the genetic reservoir for adaptation whereby they turn into adaptable variables, that is, genes subject to substantial selection (31). The essential role of gene sharing via horizontal gene transfer at the earliest stages in the evolution of life is thought of as a major factor underlying the universality of the translation system and the genetic code across all life forms (65). Strictly speaking, the transition from an ensemble of molecules to an ensemble of organisms could correspond to the emergence of protocells that lacked genomes but nevertheless displayed collective behavior and were subject to primitive form of selection for persistence (13). The origin of genomes would be a later event that kicked off natural selection. However, under the phenomenological approach adopted here, equation (7.3) covers both these stages.

The subsequent MTE, such as the origin of eukaryotic cells as a result of symbiosis between archaea and bacteria, or the origin of multicellularity, or of sociality, in principle, follow the same scheme: one has to switch between two alternative (or dual) descriptions of the same system, that is, the grand potentials in the dual descriptions should be equal at the MTE point, similar to Eq. (7.3). Here we only illustrated how the phase transition associated with the origin of life could be modeled phenomenologically and argue that essentially the same phenomenological approach would generally apply to the other MTEs.

## 8. Discussion

Since its emergence in the Big Bang about 13.8 billion years ago, our Universe has been evolving in the overall direction of increasing entropy, according to the second law of thermodynamics. Locally, however, numerous structures emerge that are characterized by readily discernible (even if not necessarily easily described formally) order and complexity. The dynamics of such structures was addressed by non-equilibrium thermodynamics (66), but traditionally has not been described as a process involving learning or selection although some attempts in this direction have been made (67, 68). However, when learning is conceived of as a universal process, under the "world as a neural network" concept (60), there is no reason not to consider all evolutionary processes in the universe within the framework of the theory of learning. Under this perspective, all systems that evolve complexity, from atoms to molecules to organisms to galaxies, learn how to predict changes in their environment with increasing accuracy and those that succeed in such prediction are selected for their stability, ability to persist and, in some cases, to propagate. During this dynamics, learning systems that evolve multiple levels of trainable variables that substantially differ in their rates of change outcompete those without such scale separation. More specifically, as argued in the accompanying paper, scale separation is considered to be a pre-requisite for the origin of life (13).

Here we combine thermodynamics of learning (12) with the theory of evolution as learning (13), in an attempt to construct a formal framework for a phenomenological description of evolution. In doing so, we continue along the lines of the previous efforts on establishing the correspondence between thermodynamics and evolution (8, 9). However, we take a more consistent statistical approach, starting from the maximum entropy principle, and introducing the principal concepts of thermodynamics and learning, which find natural counterparts in evolutionary population genetics, and we believe are indispensable for understanding evolution. The key idea of our theoretical construction is the interplay between the entropy increase in the environment dictated by the second law of thermodynamics and the entropy decrease in evolving systems (such as organisms or populations) dictated by the second law of learning (12). Under the statistical description of evolution, Malthusian fitness is naturally defined as the negative exponent of the average loss function, establishing the direct connection between the processes of evolution and learning. Further, biological temperature is defined as the inverse of the Lagrange multiplier that constrains the average loss function. This interpretation of biological temperature is related to that given by Sella and Hirsh (8), where biological temperature was represented by the inverse of the effective population size, but is more general, reflecting the degree of stochasticity in the evolutionary process, which depends not only on the effective population size, but also on other factors, in particular, interaction of organisms with the environment.

Within our theoretical framework, adaptive evolution involves primarily organisms learning to predict their environment, and accordingly, the entropy of the environment with respect to the organism is one of the key determinants of evolution. For illustration, we consider a specific phenomenological model, in which the rate of adaptive evolution reflected in the value of the loss function depends exponentially on the number of adaptable variables and also shows a power law dependence on the entropy of the environment. The number of adaptable variables, or in biological terms, the number of genes or sites that are available for positive selection in a given evolving population at a given time, is itself proportional to the entropy of the environment and to the log of the effective population size. Thus, high entropy environments promote adaptation, and then, success breeds success, that is, adaptation is most effective in large populations. These predictions of the phenomenological theory are at least qualitatively compatible with the available data and are quantitatively testable as well.

Modern evolutionary theory includes an elaborate mathematical description of microevolution (7, 69), but, to our knowledge, there is no coherent theoretical representation of MTE. Here we address this problem directly and propose a theoretical framework for MTE analysis, in which the MTE are treated as phase transitions, in the technical, physical sense. Specifically, a transition is the point where two distinct grand potentials, those characterizing units at different levels, such as molecules vs cells (organisms) in the case of the origin of life, become equal or dual. Put another way, the transition is from an ensemble of entities at a lower level of organization (for example, molecules) to an ensemble of higher level entities (for example, organisms). At the new level of organization, the lower level units display collective behavior and the corresponding phenomenological description applies. This formalism entails the existence of a critical (biological) temperature for the transition: the evolving systems have to be sufficiently robust and resistant to fluctuations for the transition to occur. Notably, this theory implies the existence of two distinct types of phase transitions in evolution: apart from MTE, each event of an adaptive mutation fixation also is a bona fide transition albeit on a much smaller scale.

The phenomenological theory of evolution outlined here is highly abstract and requires extensive further elaboration, specification and, most importantly, validation with empirical data; we indicate several specific predictions for which such validation appears to be straightforward. Nevertheless, even in this general form, the theory achieves the crucial goal of merging learning and thermodynamics into a single, coherent framework for modeling biological evolution. Therefore, we hope that this work will stimulate the development of new directions in the study of the origin of life and other MTE. By incorporating biological evolution into the framework of learning processes, this theory implies that the emergence of complexity commensurate with life is an inherent feature of learning that occurs throughout the history of our universe. Thus, although the origin of life is likely to be rare due to the multiple constraints on the learning/evolutionary processes leading to such an event (including the requirement for the essential chemicals, concentration mechanisms and more), it might not be an extremely improbable, lucky accident, but rather, a manifestation of a general evolutionary trend in a universe modeled as a learning system (13, 60).

**Author contributions**

V.V., Y.I.W., M.I.K. and E.V.K. conceptualized the project; V.V. developed the mathematical framework with contributions from M.I.K.; V.V. and E.V.K. wrote the manuscript with contributions from all authors.

**Acknowledgements**
V.V. was supported in part by the Foundational Questions Institute (FQXi) and the Oak Ridge Institute for Science and Education (ORISE). Y.I.W. and E.V.K. are supported by the Intramural Research Program of the National Institutes of Health of the USA.